\documentstyle[12pt,epsf]{article}

\begin{document}
\begin{flushright}
SLAC-PUB-7604\\
July 1997 \\
\end{flushright}
\vfill
\begin{center}
{\LARGE {Exclusive Photon-Photon Processes}
\footnote{\baselineskip=13pt Work supported by the Department
of Energy, contract DE--AC03--76SF00515.}}

\vspace{15mm}
{\bf S. J. Brodsky}\\
\vspace{5mm}
{\em Stanford Linear Accelerator Center, \\
Stanford University, Stanford, California 94309}
\end{center}
\vfill
\begin{center}
Invited Talk given at the\\
International Conference on the Structure and the Interactions\\
of the Photon (Photon97) including the \\
11th International Workshop on Photon-Photon Collisions\\
Egmond ann Zee, The Netherlands \\
May 10--15, 1997 
\end{center}

\vfill
\newpage

\begin{center}
Abstract
\end{center}
 
Exclusive $\gamma\gamma \rightarrow$ hadron pairs are among the most
fundamental processes in QCD, providing a detailed examination of
Compton scattering in the crossed channel.  In the high momentum
transfer domain $(s,t, \mbox{large},\theta_{cm}$ for $t/s$ fixed),
these processes can be computed from first principles in QCD, yielding
important information on the nature of the QCD coupling $\alpha_s$ and
the form of hadron distribution amplitudes.  Similarly, the transition
form factors $\gamma^*\gamma$, $\gamma^*\gamma \rightarrow \pi^0,
\eta^0,\eta^\prime,\eta_c \ldots$ provide rigorous tests of QCD and
definitive determinations of the meson distribution amplitudes
$\phi_H(x,Q)$.  We show that the assumption of a frozen coupling at low
momentum transfers can explain the observed scaling of two-photon
exclusive processes.

\bigskip \bigskip

\renewcommand{\bar}[1]{\overline{#1}}
\newcommand{\etal} {{\em et al.}}
\newcommand{\ie}   {{\em i.e.}}
\newcommand{\eg}   {{\em e.g.}}
\newcommand{\M}    {{\cal M}}

\section{Introduction}

Exclusive two-photon processes provide highly valuable probes of
coherent effects in quantum chromodynamics.  For example, in the case
of exclusive final states at high momentum transfer and fixed
$\theta_{cm}$ such as $\gamma \gamma \rightarrow p \bar p $ or meson
pairs, photon-photon collisions provide a timelike microscope for
testing fundamental scaling laws of PQCD and for measuring distribution
amplitudes, the fundamental wavefunctions of
hadrons.~\cite{BrodskyLepage}  At very high energies $s >> -t$ ,
diffractive processes such as $\gamma \gamma \to$ neutral vector (or
pseudoscalar) meson pairs with real or virtual photons can test the QCD
Pomeron (or the $C= -1$ exchange Odderon)  in a detailed way utilizing
the simplest possible initial state.~\cite{Ginzburg}  In the case of
low momentum transfer processes, the comparison of the two-photon decay
width for a given $C=+$ resonance with its inferred two-gluon width
provides an indirect discovery tool for gluonium. As discussed at this
conference by H. Paar, \cite{Paar} CLEO has reported a very small upper
limit for the coupling $\Gamma (\gamma\gamma \rightarrow f_J^0(1220$)
due to the absence of a signal for $K_s K_s$ decays, whereas a large
$gg\rightarrow f_J^0(1220)$ coupling  is inferred from Mark III and BES
observations of $J/\psi \rightarrow \gamma f_J^0$ decays. Using
Chanowitz's ``stickiness''  criteria, \cite{Chanowitz} this points to a
gluonium interpretation of the $f_J^0$.

Traditionally,  $\gamma\gamma$ data has come from the annihilation of
Weis\"acker--Williams effective photons emitted in  $e^-e^\pm$
collisions. Data for $\gamma\gamma \rightarrow$ hadrons from $e p
\rightarrow e^\prime p^\prime R^0$ events at HERA has also now become
available.  The HERA diffractive events will allow studies of photon
and pomeron interference effects in hadron-induced amplitudes.  As
emphasized by Klein, \cite{Klein} nuclear-coherent $\gamma\gamma
\rightarrow$ hadrons reactions can be observed in heavy-ion collisions
at RHIC or the LHC, \eg\ $Z_1 Z_2\rightarrow Z_1 Z_2 \pi^+\pi^-$.
Eventually $\gamma\gamma$ collisions  will be studied at TeV energies
with back-scattered laser beams, allowing critical probes of Standard
Model and supersymmetric processes with polarized photons in exclusive
channels such as Higgs production $\gamma \gamma \rightarrow W^+ W^-$,
and $\gamma \gamma \rightarrow W^+ W^- W^+ W^-$.~\cite{ZerwasBrodsky}

\section{Hard Exclusive Two-Photon Reactions}

Exclusive two-photon processes such as $\gamma \gamma \to $ hadron
pairs and the transition form factor $\gamma^* \gamma \to $ neutral
mesons   play a unique role  in testing quantum chromodynamics because
of the simplicity of the initial state. \cite{BrodskyLepage}  At large
momentum transfer the direct point-like coupling of the photon
dominates at leading twist, leading to highly specific predictions
which depend on the shape and normalization of the hadron distribution
amplitudes $\phi_H(x_i,Q)$ the basic valence bound state wavefunctions.
The most recent exclusive two-photon process data from CLEO
\cite{Dominick} provides stringent tests of these fundamental QCD
predictions.

Exclusive processes are particularly challenging to compute in QCD because of
their sensitivity to the unknown non-perturbative bound state dynamics of the
hadrons. However, in some important cases, the leading power-law behavior of an
exclusive amplitude at large momentum transfer can be computed rigorously via a
factorization theorem which separates the soft and hard dynamics. The
key ingredient is the factorization of the hadronic amplitude at
leading twist. As in the case of inclusive reactions, factorization
theorems for exclusive processes \cite{BrodskyLepage,EfremovRad,BLReview} allow
the analytic separation of the perturbatively-calculable short-distance
contributions from the long-distance non-perturbative dynamics
associated with hadronic binding.  
For example, the
amplitude $\gamma\gamma \rightarrow \pi^+\pi^-$ factorizes in the form
\begin{equation} 
\M_{\gamma\gamma \rightarrow \pi^+\pi^-} = \int^1_0 dx \int^1_0 dy\,
\phi_\pi(x,\widetilde Q)\, T_H(x,y,\widetilde Q)\,
\phi_\pi(y,\widetilde Q) 
\end{equation} 
where $\phi_\pi(x,\widetilde Q)$ is in the pion distribution amplitude
and contains all of the soft, nonperturbative dynamics of the pion
$q\bar q$ wavefunction integrated in relative transverse momentum up to
the separation scale $k_\perp^2 < \widetilde Q^2$, and $T_H$ is the
quark/gluon hard scattering amplitude for $\gamma\gamma \rightarrow
(q\bar q)(q\bar q)$ where the outgoing quarks are taken collinear with
their respective pion parent.  To lowest order in $\alpha_s$, the hard
scattering amplitude is linear in $\alpha_s$.  The most convenient
definition of the coupling is the effective charge $\alpha_V(Q^2)$,
defined from the potential for the scattering of two infinitely heavy
test charges, in analogy to the definition of the QED running coupling.
Another possible choice is the effective charge $\alpha_R(s)$, defined
from the QCD correction to the annihilation cross section: $R_{e^+e^-
\to {\rm hadrons}}(s) \equiv R_0 (1 + \alpha_R(s)/\pi).$ One can relate
$\alpha_V$ and $\alpha_R$ to $\alpha_{\bar{MS}}$ to NNLO using
commensurate scale relations \cite{CSR}.

The contributions from non-valence Fock states and the correction from
neglecting the transverse momentum in the subprocess amplitude from the
non-perturbative region are higher twist, {\em i.e.}, power-law
suppressed. The transverse momenta in the perturbative domain lead to
the evolution of the distribution amplitude and to
next-to-leading-order (NLO) corrections in $\alpha_s$.  The
contribution from the endpoint regions of integration, $x \sim 1$ and
$y \sim 1,$ are power-law and Sudakov suppressed and thus can only
contribute corrections at higher order in $1/Q$. \cite{BrodskyLepage}

The distribution amplitude $\phi(x,\widetilde Q)$ is boost and gauge
invariant and evolves in $\ln \widetilde Q$ through an evolution equation
\cite{BrodskyLepage}.  It can be computed from the integral over
transverse momenta of the renormalized hadron valence wavefunction in
the light-cone gauge at fixed light-cone time \cite{BrodskyLepage}:
\begin{equation}
\phi(x,\widetilde Q) = \int d^2\vec{k_\perp}\thinspace
\theta \left({\widetilde Q}^2 - {\vec{k_\perp}^2\over x(1-x)}\right)
\psi^{(\widetilde Q)}(x,\vec{k_\perp}).
\label{quarkdistamp}
\end{equation}
A physical amplitude must be independent of the separation scale
$\widetilde Q.$ The natural variable in which to make this separation
is the light-cone energy, or equivalently the invariant mass ${\cal
M}^2 ={\vec{k_\perp}^2/ x(1-x)}$, of the off-shell partonic system
\cite{JIPang,BrodskyLepage}. Any residual dependence on the choice of
$\widetilde Q$ for the distribution amplitude will be compensated by a
corresponding dependence of the NLO correction in $T_H.$  In general,
the NLO prediction for exclusive amplitude depends strongly on the form
of the pion distribution amplitude as well as the choice of
renormalization scale $\mu$ and scheme.
 
The QCD coupling is typically evaluated at quite low scales in
exclusive processes since the momentum transfers has to be divided
among several constituents.  In the BLM procedure, the scale of the
coupling is evaluated by absorbing all vacuum polarization corrections
with the scale of the coupling or by taking the experimental value
integrating over the gluon virtuality.  Thus, in the case of the
(timelike) pion form factor the relevant scale is of order $Q^{*2} \sim
e^{-3}\M^2_{\pi\pi^-} \cong \frac{1}{20}\, \M^2_{\pi^+\pi^-}$ assuming
the asymptotic form of the pion distribution amplitude $\phi^{\rm
asympt}_\pi = \sqrt 3\, f_\pi\, x(1-x)$.  At such low scales, it is
likely that the coupling is frozen or relatively slow varying.

In the BLM procedure, the renormalization scales are chosen such that
all vacuum polarization effects from the QCD $\beta$ function are
re-summed into the running couplings.  The coefficients of the
perturbative series are thus identical to the perturbative coefficients
of the corresponding conformally invariant theory with $\beta=0.$ The
BLM method has the important advantage of ``pre-summing" the large and
strongly divergent terms in the PQCD series which grow as $n! 
(\alpha_s \beta_0 )^n$, {\em i.e.}, the infrared renormalons associated
with coupling constant renormalization \cite{Mueller,BallBenekeBraun}.
Furthermore, the renormalization scales $Q^*$ in the BLM method are
physical in the sense that they reflect the mean virtuality of the
gluon propagators \cite{BallBenekeBraun,BLM,LepageMackenzie,Neubert}. 
In fact, in the $\alpha_V(Q)$ scheme, where the QCD coupling is defined
from the heavy quark potential, the renormalization scale is by
definition the momentum transfer caused by the gluon. Because the
renormalization scale is small in the exclusive $\gamma \gamma$
processes discussed here, we will argue that the effective coupling is
nearly constant, thus accounting for the nominal scaling behavior of
the data \cite{JiSillLombard,JiAmiri}.

The heavy-quark potential $V(Q^2)$ can be identified via the
two-particle-irreducible scattering amplitude of test charges, {\em
i.e.}, the scattering of an in\-fi\-nite\-ly heavy quark and antiquark at
momentum transfer $t = -Q^2.$ The relation
\begin{equation}
V(Q^2) = -  {4 \pi C_F \alpha_V(Q^2)\over Q^2},
\end{equation}
with $C_F=(N_C^2-1)/2 N_C=4/3$, then defines the effective charge
$\alpha_V(Q)$.  This coupling provides a physically-based alternative
to the usual ${\overline {MS}}$ scheme.    As in the corresponding case of Abelian QED, the
scale $Q$ of the coupling $\alpha_V(Q)$ is identified with the
exchanged momentum. The scale-fixed relation between $\alpha_V$ and the
conventional $\overline {MS}$ coupling is
\begin{equation}
\alpha_V(Q) = \alpha_{\overline {MS}}(e^{-5/6} Q) \left(1 -
\frac{2C_A}{3} {\alpha_{\overline {MS}}\over\pi} + \cdots\right),
\label{alpmsbar}
\end {equation}
above or below any quark mass threshold.  The factor $e^{-5/6} \simeq
0.4346$ is the ratio of commensurate scales between the two schemes to
this order.  It arises because of the conventions used in defining the
modified minimal subtraction scheme. The scale in the $\overline {MS}$
scheme is thus a factor $\sim 0.4$ smaller than the physical scale. 
The coefficient $2C_A/3$ in the NLO term is a feature of the
non-Abelian couplings of QCD; the same coefficient would occur even if
the theory were conformally invariant with $\beta_0=0.$ 
Recent lattice calculations
have provided strong constraints on the normalization and shape of
$\alpha_V(Q^2)$. \cite{Davies} The $J/\psi$ and $\Upsilon$ spectra
have been used to determine the normalization:
\begin{equation}
\alpha_V^{(3)}(8.2~{\rm GeV}) = 0.196(3),
\label{alpv8.2}
\end{equation}
where the effective number of light flavors is $n_f = 3$. The
corresponding modified minimal subtraction coupling evolved to the $Z$
mass using Eq. (\ref{alpmsbar}) is given by
\begin{equation}
\alpha_{\overline{MS}}^{(5)}(M_Z) = 0.115(2).
\label{alpmsbarmz}
\end{equation}
This value is consistent with the world average of 0.117(5), but is
significantly more precise. These results are valid up to NLO.

Ji, Pang, Robertson, and I \cite{BJPR} have recently analyzed the pion
transition form factor $F^{\gamma^*\gamma} \rightarrow \pi^0$ obtained
from $e\gamma \rightarrow e^\prime \pi^0$, the timelike pion form
obtained from $e^+e^- \rightarrow \pi^+\pi$, and the $\gamma\gamma
\rightarrow \pi^+\pi^-$ processes, all at NLO in $\alpha_V$. The
assumption of a nearly constant coupling in the hard scattering
amplitude at low scales provides an explanation for the
phenomenological success of dimensional counting rules for exclusive
processes; \ie, the power-law fall-off follows the nominal scaling of
the hard scattering amplitude $\M_{\rm had} \sim T_H\sim [p_T]^{4-n}$
where $n$ is in the total number of incident and final fields entering
$T_H$. The transition form factor has now been measured up to $Q^2 < 8
$ GeV$^2$ in the tagged two-photon collisions $e \gamma \to e' \pi^0$
by the CLEO and CELLO collaborations.  In this case the amplitude has
the factorized form
\begin{equation}
F_{\gamma M}(Q^2)= {4 \over \sqrt 3}\int^1_0 dx \phi_M(x,Q^2)
T^H_{\gamma \to M}(x,Q^2) ,
\label{transitionformfactor}
\end{equation}
where the hard scattering amplitude for $\gamma \gamma^* \to q \bar q$
is
\begin{equation}
T^H_{\gamma M}(x,Q^2) = {1\over (1-x) Q^2}\left(1 +
{\cal O}(\alpha_s)\right).
\label{transitionhardscattering}
\end{equation}
The leading QCD corrections have been computed by Braaten
\cite{Braaten}; however, the NLO corrections are necessary to fix the
BLM scale at LO.  Thus it is not yet possible to rigorously determine
the BLM scale for this quantity.  We shall here assume that this scale
is the same as that occurring in the prediction for $F_\pi$.  For the
asymptotic distribution amplitude we thus predict
\begin{equation}
Q^2 F_{\gamma \pi}(Q^2)= 2 f_\pi \left(1 - {5\over3}
{\alpha_V(Q^*)\over \pi}\right).
\label{qsquaretransition}
\end{equation}
As we shall see, given the phenomenological form of $\alpha_V$ we
employ (discussed below), this result is not terribly sensitive to the
precise value of the scale.

An important prediction resulting from the factorized form of these
results is that the normalization of the ratio
\begin{eqnarray}
R_\pi(Q^2) &\equiv & \frac{F_\pi (Q^2)}{4 \pi Q^2 |F_{\pi
\gamma}(Q^2)|^2}
\label{Rpidef}\\
&=& \alpha_{\overline{MS}}(e^{-14/6}Q)\left(1-0.56
{\alpha_{\overline{MS}} \over\pi} \right)
\label{Rpistart}\\
&=& \alpha_V(e^{-3/2}Q)\left(1+1.43 {\alpha_V\over\pi} \right)
\label{RpiV}\\
&=& \alpha_R(e^{5/12-{2\zeta_3}}Q) \left(1-
0.65 {\alpha_R\over\pi} \right)
\label{Rpiend}
\end{eqnarray}
is formally independent of the form of the pion distribution amplitude.
The $\alpha_{\overline{MS}}$ correction follows from combined
references \cite{Braaten,DittesRadyushkin,Field}.  The next-to-leading
correction given here assumes the asymptotic distribution amplitude.

We emphasize that when we relate $R_\pi$ to $\alpha_V$ we relate
observable to observable and thus there is no scheme ambiguity. 
Furthermore, effective charges such as $\alpha_V$  are defined from
physical observables and thus must be finite even at low momenta. A
number of proposals have been suggested for the form of the QCD
coupling in the low-momentum regime.  For example, Petronzio and Parisi
\cite{PetronzioParisi} have argued that the coupling must freeze at low
momentum transfer in order that perturbative QCD loop integrations be
well defined.  Mattingly and Stevenson \cite{MattinglyStevenson} have
incorporated such behavior into their parameterizations of $\alpha_R$
at low scales.  Gribov \cite{Gribov} has presented novel dynamical
arguments related to the nature of confinement for a fixed coupling at
low scales.  Zerwas \cite{Zerwas} has noted the heavy quark potential
must saturate to a Yukawa form since the light-quark production
processes will screen the linear confining potential at large
distances.  Cornwall \cite{Cornwall} and others
\cite{DonnachieLandshoff,GayDucatietal} have argued that the gluon
propagator will acquire an effective gluon mass $m_g$ from
non-perturbative dynamics, which again will regulate the form of the
effective couplings at low momentum. We shall adopt the simple
parameterization
\begin{equation}
\alpha_V(Q) = {4 \pi \over {\beta_0 \ln \left({{Q^2 + 4m_g^2}\over
\Lambda^2_V}\right)}} ,
\label{frozencoupling}
\end{equation}
which effectively freezes the $\alpha_V$ effective charge to a finite
value for $Q^2 \leq 4m_g^2.$

\begin{figure}[htb]
\begin{center}
\hbox to 4in{\input{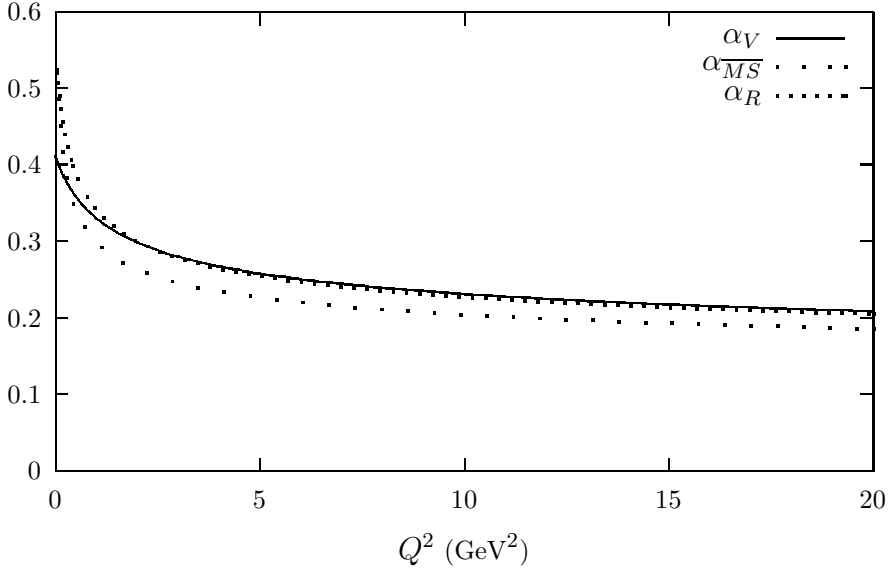}}
\end{center}
\caption[*]{The coupling function $\alpha_V(Q^2)$ as given in
Eq. (\ref{frozencoupling}).  Also shown are the corresponding
predictions for $\alpha_{\overline{MS}}$ and $\alpha_R$ following from
the NLO commensurate scale relations.}
\label{couplings}
\end{figure}

We can use the non-relativistic heavy quark lattice results
\cite{Davies,Sloan} to fix the parameters.  A fit to the lattice data
of the above parameterization gives $\Lambda_V = 0.16$ GeV if we use
the well-known momentum-dependent $n_f$ \cite{ShirkovMikhailov}.
Furthermore, the value $m^2_g=0.19$ GeV$^2$ gives consistency with the
frozen value of $\alpha_R$ advocated by Mattingly and Stevenson
\cite{MattinglyStevenson}.  Their parameterization implies the
approximate constraint $\alpha_R(Q)/\pi \simeq 0.27$ for $Q= \sqrt s <
0.3$ GeV, which leads to $\alpha_V(0.5~{\rm GeV}) \simeq 0.37$ using
the NLO commensurate scale relation between $\alpha_V$ and $\alpha_R$.
The resulting form for $\alpha_V$ is shown in Fig. \ref{couplings}. The
corresponding predictions for $\alpha_R$ and $\alpha_{\overline {MS}}$
using the CSRs at NLO are also shown.  Note that for low $Q^2$ the
couplings, although frozen, are large.  Thus the NLO and higher-order
terms in the CSRs are large, and inverting them perturbatively to NLO
does not give accurate results at low scales.  In addition,
higher-twist contributions to $\alpha_V$ and $\alpha_R$, which are not
reflected in the CSR relating them, may be expected to be important for
low $Q^2$ \cite{Braun}.

It is clear that exclusive processes such as the photon to pion
transition form factors can provide a valuable window for determining
the magnitude and the shape of the effective charges at quite low
momentum transfers.  In particular, we can check consistency with the
$\alpha_V$ prediction from lattice gauge theory.  A complimentary
method for determining $\alpha_V$ at low momentum is to use the angular
anisotropy of $e^+ e^- \to Q \overline Q$ at the heavy quark thresholds
\cite{BrodskyKuhnHoangTuebner}. It should be emphasized that the
parameterization (\ref{frozencoupling}) is just an approximate form.
The actual behavior of $\alpha_V(Q^2)$ at low $Q^2$ is one of the key
uncertainties in QCD phenomenology.

As we have emphasized, exclusive processes are sensitive to the
magnitude and shape of the QCD couplings at quite low momentum
transfer: $Q_V^{*2} \simeq e^{-3} Q^2 \simeq Q^2/20$ and $Q_R^{*2}
\simeq Q^2/50$ \cite{LewellynIsgur}.  The fact that the data for
exclusive processes such as form factors, two photon processes such as
$\gamma \gamma \to \pi^+ \pi^-,$ and photoproduction at fixed
$\theta_{c.m.}$ are consistent with the nominal scaling of the
leading-twist QCD predictions (dimensional counting) at momentum
transfers $Q$ up to the order of a few GeV can be immediately
understood if the effective charges $\alpha_V$ and $\alpha_R$ are
slowly varying at low momentum.  The scaling of the exclusive amplitude
then follows that of the subprocess amplitude $T_H$ with effectively
fixed coupling. Note also that the Sudakov effect of the end point
region is the exponential of a double log series if the coupling is
frozen, and thus is strong.

\begin{figure}[htb]
\begin{center}
\hbox to 4in{
\setlength{\unitlength}{0.240900pt}
\ifx\plotpoint\undefined\newsavebox{\plotpoint}\fi
\sbox{\plotpoint}{\rule[-0.200pt]{0.400pt}{0.400pt}}%
\begin{picture}(1500,900)(0,0)
\font\gnuplot=cmr10 at 10pt
\gnuplot
\sbox{\plotpoint}{\rule[-0.200pt]{0.400pt}{0.400pt}}%
\put(219.0,134.0){\rule[-0.200pt]{4.818pt}{0.400pt}}
\put(197,134){\makebox(0,0)[r]{0}}
\put(1416.0,134.0){\rule[-0.200pt]{4.818pt}{0.400pt}}
\put(219.0,278.0){\rule[-0.200pt]{4.818pt}{0.400pt}}
\put(197,278){\makebox(0,0)[r]{0.05}}
\put(1416.0,278.0){\rule[-0.200pt]{4.818pt}{0.400pt}}
\put(219.0,422.0){\rule[-0.200pt]{4.818pt}{0.400pt}}
\put(197,422){\makebox(0,0)[r]{0.1}}
\put(1416.0,422.0){\rule[-0.200pt]{4.818pt}{0.400pt}}
\put(219.0,567.0){\rule[-0.200pt]{4.818pt}{0.400pt}}
\put(197,567){\makebox(0,0)[r]{0.15}}
\put(1416.0,567.0){\rule[-0.200pt]{4.818pt}{0.400pt}}
\put(219.0,711.0){\rule[-0.200pt]{4.818pt}{0.400pt}}
\put(197,711){\makebox(0,0)[r]{0.2}}
\put(1416.0,711.0){\rule[-0.200pt]{4.818pt}{0.400pt}}
\put(219.0,855.0){\rule[-0.200pt]{4.818pt}{0.400pt}}
\put(197,855){\makebox(0,0)[r]{0.25}}
\put(1416.0,855.0){\rule[-0.200pt]{4.818pt}{0.400pt}}
\put(219.0,134.0){\rule[-0.200pt]{0.400pt}{4.818pt}}
\put(219,89){\makebox(0,0){0}}
\put(219.0,835.0){\rule[-0.200pt]{0.400pt}{4.818pt}}
\put(462.0,134.0){\rule[-0.200pt]{0.400pt}{4.818pt}}
\put(462,89){\makebox(0,0){2}}
\put(462.0,835.0){\rule[-0.200pt]{0.400pt}{4.818pt}}
\put(706.0,134.0){\rule[-0.200pt]{0.400pt}{4.818pt}}
\put(706,89){\makebox(0,0){4}}
\put(706.0,835.0){\rule[-0.200pt]{0.400pt}{4.818pt}}
\put(949.0,134.0){\rule[-0.200pt]{0.400pt}{4.818pt}}
\put(949,89){\makebox(0,0){6}}
\put(949.0,835.0){\rule[-0.200pt]{0.400pt}{4.818pt}}
\put(1193.0,134.0){\rule[-0.200pt]{0.400pt}{4.818pt}}
\put(1193,89){\makebox(0,0){8}}
\put(1193.0,835.0){\rule[-0.200pt]{0.400pt}{4.818pt}}
\put(1436.0,134.0){\rule[-0.200pt]{0.400pt}{4.818pt}}
\put(1436,89){\makebox(0,0){10}}
\put(1436.0,835.0){\rule[-0.200pt]{0.400pt}{4.818pt}}
\put(219.0,134.0){\rule[-0.200pt]{293.175pt}{0.400pt}}
\put(1436.0,134.0){\rule[-0.200pt]{0.400pt}{173.689pt}}
\put(219.0,855.0){\rule[-0.200pt]{293.175pt}{0.400pt}}
\put(-22,494){\makebox(0,0){\shortstack{$Q^2 F_{\gamma\pi} (Q^2)$\\ \\ (GeV)}}}
\put(827,10){\makebox(0,0){$Q^2$ (GeV$^2$)}}
\put(219.0,134.0){\rule[-0.200pt]{0.400pt}{173.689pt}}
\put(419,483){\circle*{12}}
\put(450,471){\circle*{12}}
\put(475,532){\circle*{12}}
\put(499,500){\circle*{12}}
\put(523,523){\circle*{12}}
\put(548,569){\circle*{12}}
\put(577,529){\circle*{12}}
\put(619,552){\circle*{12}}
\put(674,515){\circle*{12}}
\put(735,520){\circle*{12}}
\put(796,578){\circle*{12}}
\put(857,552){\circle*{12}}
\put(918,581){\circle*{12}}
\put(1006,561){\circle*{12}}
\put(1180,616){\circle*{12}}
\put(419.0,451.0){\rule[-0.200pt]{0.400pt}{15.418pt}}
\put(409.0,451.0){\rule[-0.200pt]{4.818pt}{0.400pt}}
\put(409.0,515.0){\rule[-0.200pt]{4.818pt}{0.400pt}}
\put(450.0,443.0){\rule[-0.200pt]{0.400pt}{13.731pt}}
\put(440.0,443.0){\rule[-0.200pt]{4.818pt}{0.400pt}}
\put(440.0,500.0){\rule[-0.200pt]{4.818pt}{0.400pt}}
\put(475.0,500.0){\rule[-0.200pt]{0.400pt}{15.418pt}}
\put(465.0,500.0){\rule[-0.200pt]{4.818pt}{0.400pt}}
\put(465.0,564.0){\rule[-0.200pt]{4.818pt}{0.400pt}}
\put(499.0,466.0){\rule[-0.200pt]{0.400pt}{16.622pt}}
\put(489.0,466.0){\rule[-0.200pt]{4.818pt}{0.400pt}}
\put(489.0,535.0){\rule[-0.200pt]{4.818pt}{0.400pt}}
\put(523.0,486.0){\rule[-0.200pt]{0.400pt}{18.067pt}}
\put(513.0,486.0){\rule[-0.200pt]{4.818pt}{0.400pt}}
\put(513.0,561.0){\rule[-0.200pt]{4.818pt}{0.400pt}}
\put(548.0,526.0){\rule[-0.200pt]{0.400pt}{20.958pt}}
\put(538.0,526.0){\rule[-0.200pt]{4.818pt}{0.400pt}}
\put(538.0,613.0){\rule[-0.200pt]{4.818pt}{0.400pt}}
\put(577.0,486.0){\rule[-0.200pt]{0.400pt}{20.717pt}}
\put(567.0,486.0){\rule[-0.200pt]{4.818pt}{0.400pt}}
\put(567.0,572.0){\rule[-0.200pt]{4.818pt}{0.400pt}}
\put(619.0,506.0){\rule[-0.200pt]{0.400pt}{22.163pt}}
\put(609.0,506.0){\rule[-0.200pt]{4.818pt}{0.400pt}}
\put(609.0,598.0){\rule[-0.200pt]{4.818pt}{0.400pt}}
\put(674.0,466.0){\rule[-0.200pt]{0.400pt}{23.608pt}}
\put(664.0,466.0){\rule[-0.200pt]{4.818pt}{0.400pt}}
\put(664.0,564.0){\rule[-0.200pt]{4.818pt}{0.400pt}}
\put(735.0,469.0){\rule[-0.200pt]{0.400pt}{24.813pt}}
\put(725.0,469.0){\rule[-0.200pt]{4.818pt}{0.400pt}}
\put(725.0,572.0){\rule[-0.200pt]{4.818pt}{0.400pt}}
\put(796.0,518.0){\rule[-0.200pt]{0.400pt}{29.149pt}}
\put(786.0,518.0){\rule[-0.200pt]{4.818pt}{0.400pt}}
\put(786.0,639.0){\rule[-0.200pt]{4.818pt}{0.400pt}}
\put(857.0,489.0){\rule[-0.200pt]{0.400pt}{30.594pt}}
\put(847.0,489.0){\rule[-0.200pt]{4.818pt}{0.400pt}}
\put(847.0,616.0){\rule[-0.200pt]{4.818pt}{0.400pt}}
\put(918.0,506.0){\rule[-0.200pt]{0.400pt}{36.135pt}}
\put(908.0,506.0){\rule[-0.200pt]{4.818pt}{0.400pt}}
\put(908.0,656.0){\rule[-0.200pt]{4.818pt}{0.400pt}}
\put(1006.0,492.0){\rule[-0.200pt]{0.400pt}{33.244pt}}
\put(996.0,492.0){\rule[-0.200pt]{4.818pt}{0.400pt}}
\put(996.0,630.0){\rule[-0.200pt]{4.818pt}{0.400pt}}
\put(1180.0,532.0){\rule[-0.200pt]{0.400pt}{40.230pt}}
\put(1170.0,532.0){\rule[-0.200pt]{4.818pt}{0.400pt}}
\put(1170.0,699.0){\rule[-0.200pt]{4.818pt}{0.400pt}}
\sbox{\plotpoint}{\rule[-0.500pt]{1.000pt}{1.000pt}}%
\put(341,670){\usebox{\plotpoint}}
\multiput(341,670)(20.756,0.000){52}{\usebox{\plotpoint}}
\put(1418,670){\usebox{\plotpoint}}
\sbox{\plotpoint}{\rule[-0.200pt]{0.400pt}{0.400pt}}%
\put(341,561){\usebox{\plotpoint}}
\put(430,560.67){\rule{5.300pt}{0.400pt}}
\multiput(430.00,560.17)(11.000,1.000){2}{\rule{2.650pt}{0.400pt}}
\put(341.0,561.0){\rule[-0.200pt]{21.440pt}{0.400pt}}
\put(587,561.67){\rule{5.300pt}{0.400pt}}
\multiput(587.00,561.17)(11.000,1.000){2}{\rule{2.650pt}{0.400pt}}
\put(452.0,562.0){\rule[-0.200pt]{32.521pt}{0.400pt}}
\put(698,562.67){\rule{5.541pt}{0.400pt}}
\multiput(698.00,562.17)(11.500,1.000){2}{\rule{2.770pt}{0.400pt}}
\put(609.0,563.0){\rule[-0.200pt]{21.440pt}{0.400pt}}
\put(810,563.67){\rule{5.300pt}{0.400pt}}
\multiput(810.00,563.17)(11.000,1.000){2}{\rule{2.650pt}{0.400pt}}
\put(721.0,564.0){\rule[-0.200pt]{21.440pt}{0.400pt}}
\put(944,564.67){\rule{5.541pt}{0.400pt}}
\multiput(944.00,564.17)(11.500,1.000){2}{\rule{2.770pt}{0.400pt}}
\put(832.0,565.0){\rule[-0.200pt]{26.981pt}{0.400pt}}
\put(1056,565.67){\rule{5.300pt}{0.400pt}}
\multiput(1056.00,565.17)(11.000,1.000){2}{\rule{2.650pt}{0.400pt}}
\put(967.0,566.0){\rule[-0.200pt]{21.440pt}{0.400pt}}
\put(1190,566.67){\rule{5.300pt}{0.400pt}}
\multiput(1190.00,566.17)(11.000,1.000){2}{\rule{2.650pt}{0.400pt}}
\put(1078.0,567.0){\rule[-0.200pt]{26.981pt}{0.400pt}}
\put(1324,567.67){\rule{5.541pt}{0.400pt}}
\multiput(1324.00,567.17)(11.500,1.000){2}{\rule{2.770pt}{0.400pt}}
\put(1212.0,568.0){\rule[-0.200pt]{26.981pt}{0.400pt}}
\put(1347.0,569.0){\rule[-0.200pt]{21.440pt}{0.400pt}}
\end{picture}}
\end{center}
\caption[*]{The $\gamma\rightarrow\pi^0$ transition form factor.  The
solid line is the full prediction including the QCD correction
[Eq. (\ref{fgampi})]; the dotted line is the LO prediction
$Q^2F_{\gamma\pi}(Q^2) = 2f_\pi$.}
\label{fgammapi}
\end{figure}
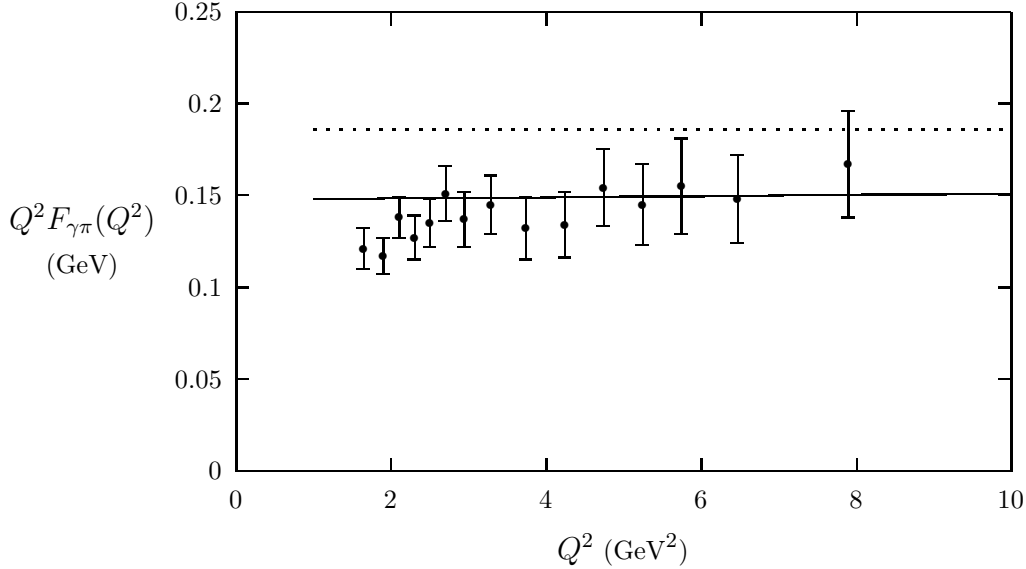

In Fig. \ref{fgammapi}, we compare the recent CLEO
data \cite{Dominick} for the photon to pion transition form factor with
the prediction
\begin{equation}
Q^2 F_{\gamma\pi}(Q^2)= 2 f_\pi \left( 1
- {5\over3} {\alpha_V(e^{-3/2} Q)\over \pi}\right).
\label{fgampi}
\end{equation}
The 
flat scaling of the $Q^2 F_{\gamma \pi}(Q^2)$ data from $Q^2 = 2$
to $Q^2 = 8$ GeV$^2$ provides an important confirmation of the
applicability of leading twist QCD to this process. The magnitude of
$Q^2 F_{\gamma \pi}(Q^2)$ is remarkably consistent with the predicted
form, assuming the asymptotic distribution amplitude and including the
LO QCD radiative correction with $\alpha_V(e^{-3/2} Q)/\pi \simeq
0.12$.  Radyushkin \cite {Radyushkin}, Ong \cite{Ong} and Kroll \cite
{Kroll} have also noted that the scaling and normalization of the
photon-to-pion transition form factor tends to favor the asymptotic
form for the pion distribution amplitude and rules out broader
distributions such as the two-humped form suggested by QCD sum rules
\cite{CZ}.  One cannot obtain a unique solution for the
non-perturbative wavefunction from the $F_{\pi\gamma}$ data alone.
However, we have the constraint that
\begin{equation}
{1\over 3}\langle {1\over 1-x}\rangle \left[ 1-{5\over
3}{\alpha_V(Q^*)\over\pi} \right]\simeq 0.8
\end{equation}
(assuming the renormalization scale we have chosen in
Eq. (\ref{qsquaretransition}) is approximately correct).  Thus one
could allow for some broadening of the distribution amplitude with a
corresponding increase in the value of $\alpha_V$ at low scales.

\begin{figure}[htb]
\begin{center}
\hbox to 4in{\input{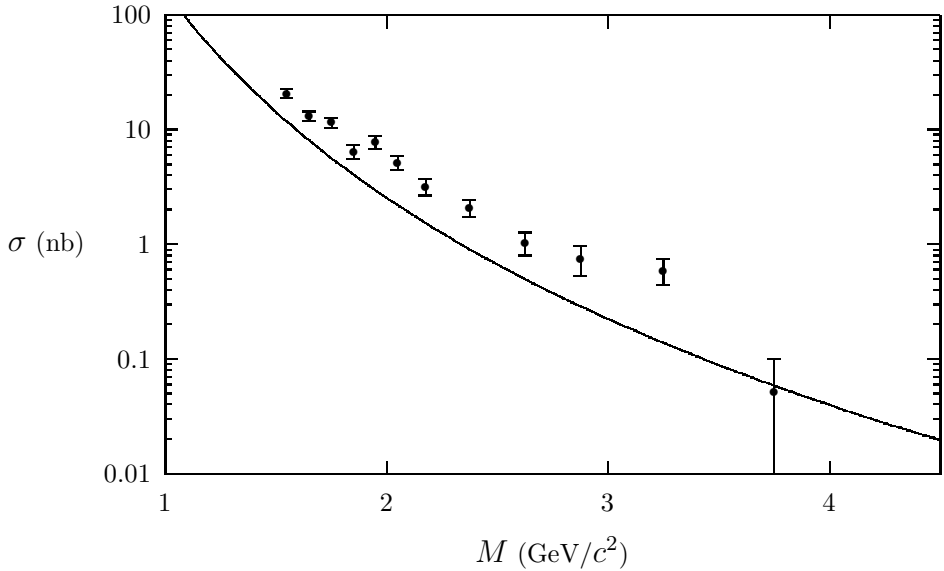}}
\end{center}
\caption[*]{Two-photon annihilation cross section $\sigma(\gamma \gamma
\to\pi^+\pi^-,K^+K^-)$ as a function of CMS energy, for
$|\cos\theta^*|<0.6$.}
\label{sigma}
\end{figure}

We have also analyzed the $\gamma\gamma \rightarrow \pi^+\pi^-, K^+
K^-$ data. These data exhibit true leading-twist scaling (Fig.
\ref{sigma}), so that one would expect this process to be a good test
of theory. One can show that to LO
\begin{equation}
{{d\sigma\over dt}\left(\gamma\gamma\rightarrow\pi^+\pi^-\right) \over
{d\sigma\over dt}\left(\gamma\gamma\rightarrow\mu^+\mu^-\right)} =
{4|F_\pi(s)|^2\over 1-\cos^4\theta_{c.m.}}
\end{equation}
in the CMS, where $dt=(s/2) d(\cos\theta_{c.m.})$ and here $F_\pi(s)$
is the {\em time-like} pion form factor.  The ratio of the time-like
to space-like pion form factor for the asymptotic distribution
amplitude is given by
\begin{equation}
{|F^{(\rm timelike)}_\pi(-Q^2)|\over F^{(\rm spacelike)}_\pi(Q^2)}
= {|\alpha_V(-Q^{*2})|\over \alpha_V(Q^{*2})}.
\label{ratio}
\end{equation}
If we simply continue Eq. (\ref{frozencoupling}) to negative values of
$Q^2$ then for $1 < Q^2 < 10$ GeV$^2$, and
hence $0.05 < Q^{*2} < 0.5$ GeV$^2$, the ratio of couplings in
Eq. (\ref{ratio}) is of order 1.5.  Of course this assumes the
analytic application of Eq. (\ref{frozencoupling}).  Thus if we assume
the asymptotic form for the distribution amplitude, then we predict
$F^{(\rm timelike)}_\pi(-Q^2) \simeq (0.3~{\rm GeV}^2)/Q^2$ and hence
\begin{equation}
{{d\sigma\over dt}\left(\gamma\gamma\rightarrow\pi^+\pi^-\right) \over
{d\sigma\over dt}\left(\gamma\gamma\rightarrow\mu^+\mu^-\right)
}\simeq {.36 \over s^2} {1\over 1-\cos^4\theta_{c.m.}}.
\label{twophotonratio}
\end{equation}
The resulting prediction for the combined cross section
$\sigma(\gamma\gamma\to\pi^+\pi^-, K^+K^-)$\footnote{The contribution
from kaons is obtained at this order simply by rescaling the
prediction for pions by a factor $(f_K/f_\pi)^4\simeq 2.2$.} is shown
in Fig. \ref{sigma}, along with CLEO data \cite{Dominick}.
Considering the possible contribution of the resonance $f_2(1270)$,
the agreement is reasonable.

We also note that the normalization of $\alpha_V$ could be larger at
low momentum than our estimate. This would also imply a broadening of
the pion distribution amplitude compared to its asymptotic form since
one needs to raise the expectation value of $1/(1-x)$ in order to
maintain consistency with the magnitude of the $F_{\gamma \pi}(Q^2)$
data. A full analysis will then also require consideration of the
breaking of scaling from the evolution of the distribution amplitude.
In any case, we find no compelling argument for significant
higher-twist contributions in the few GeV regime from the hard
scattering amplitude or the endpoint regions, since such corrections
violate the observed scaling behavior of the data.

The analysis we have presented here suggests a systematic program for
estimating exclusive amplitudes in QCD (including exclusive $B$-decays)
which involve hard scattering.  The central input is $\alpha_V(0)$, or
\begin{equation}
\overline{\alpha_V} = {1\over{Q_0^2}}\int_0^{Q_0^2}d{Q^\prime}^2
\alpha_V({Q^\prime}^2),\;\; Q_0^2 \leq 1\;\;{\rm GeV}^2,
\label{old21}
\end{equation}
which largely controls the magnitude of the underlying quark-gluon
subprocesses for hard processes in the few-GeV region.  In this work,
the mean coupling value for $Q_0^2 \simeq 0.5$ GeV$^2$ is
$\overline{\alpha_V} \simeq 0.38.$ The main focus will then be to
determine the shapes and normalization of the process-independent meson
and baryon distribution amplitudes.

\section{Conclusions}

The leading-twist scaling of the observed cross sections for exclusive
two-photon processes and other fixed $\theta_{cm}$ reactions can be
understood if the effective coupling $\alpha_V(Q^*)$ is approximately
constant in the domain of $Q^*$ relevant to the underlying hard
scattering amplitudes.  In addition, the Sudakov suppression of the
long-distance contributions is strengthened if the coupling is frozen
because of the exponentiation of a double log series.    We have also
found that the commensurate scale relation connecting the heavy quark
potential, as determined from lattice gauge theory, to the
photon-to-pion transition form factor is in excellent agreement with
$\gamma e \to \pi^0 e$ data assuming that the pion distribution
amplitude is close to its asymptotic form $\sqrt{3}f_\pi x(1-x)$.  We
also reproduce the scaling and approximate normalization of the
$\gamma\gamma \rightarrow \pi^+\pi^-, K^+ K^-$ data at large momentum
transfer. However, the normalization of the space-like pion form factor
$F_\pi(Q^2)$ obtained from electroproduction experiments is somewhat
higher than that predicted by the corresponding commensurate scale
relation. This discrepancy may be due to systematic errors introduced
by the extrapolation of the $\gamma^* p \to \pi^+ n$ electroproduction
data to the pion pole.

\section*{Acknowledgments}

\noindent
Much of this talk is based on collaborations with  Peter Lepage, Hung
Jung Lu, Chueng Ji, Dave Robertson, and Alex Pang, and I thank them for
helpful conversations. This work is supported in part by the U.S.
Department of Energy under contract no. DE--AC03--76SF00515.


\begin{thebibliography}{99}


\bibitem{BrodskyLepage}
S. J. Brodsky and G. P. Lepage, {\em Phys. Rev.  Lett.} {\bf 53}, 545
(1979); {\em Phys. Lett.} {\bf 87B}, 359 (1979); G. P. Lepage and
S. J. Brodsky, {\em Phys. Rev.} {\bf D22}, 2157 (1980).

\bibitem{Ginzburg}
I.F. Ginzburg, D.Yu. Ivanov, and V.G. Serbo
these proceedings, and e-Print hep-ph/9508309. 

\bibitem{Paar} H. Paar, these proceedings.

\bibitem{Chanowitz} 
M. Chanowitz, {\em Nucl. Instrum. Meth.} {\bf  A355} 42, (1995),  
e-Print hep-ph/9407231, and references therein.  

\bibitem{Klein}
S.  Klein,  e-Print, nucl-th/9707008  and these proceedings.

\bibitem{ZerwasBrodsky}
S. J. Brodsky and P. M. Zerwas, {\em Nucl. Instrum. Meth.} {\bf  A355}
19, (1995),   e-Print  hep-ph/9407362, and references therein.  Se also
G. Jikia, these proceedings.


\bibitem{Dominick}
J. Dominick, {\em et al.}, Phys. Rev. {\bf D50}, 3027 (1994). See also,
J. Gronberg {\em et al.} CLNS-97-1477, (1997), e-Print hep-ex/9707031
and V. Savinov, these proceedings.

\bibitem{EfremovRad} 
A. V. Efremov and A. V. Radyushkin, {\it Theor. Math. Phys.} {\bf 42},
97 (1980).

\bibitem{BLReview}
S. J. Brodsky and G. P. Lepage, in {\em Perturbative Quantum
Chromodynamics}, A. H. Mueller, Ed.  (World Scientific, 1989).

\bibitem{CSR}
H. J. Lu and S. J. Brodsky, {\em Phys. Rev.} {\bf D48}, 3310 (1993).

\bibitem{JIPang}
C.-R. Ji, A. Pang, and A. Szczepaniak, {\em Phys. Rev.} {\bf D52},
4038 (1995).

\bibitem{Mueller}
A. H. Mueller, {\em Nucl. Phys.} {\bf B250}, 327 (1985).

\bibitem{BallBenekeBraun}
P. Ball, M. Beneke and V. M. Braun, {\em Phys. Rev.} {\bf D52}, 3929
(1995).


\bibitem{BLM}
S. J. Brodsky, G. P. Lepage, and P. B. Mackenzie, {\em Phys. Rev.}
{\bf D28}, 228 (1983).

\bibitem{LepageMackenzie}
G. P. Lepage and P. B.  Mackenzie, {\em Phys. Rev.} {\bf D48}, 2250
(1993).

\bibitem{Neubert}
M. Neubert, {\em Phys. Rev.} {\bf D51}, 5924 (1995); {\em
Phys. Rev. Lett.} {\bf 76}, 3061 (1996).

\bibitem{JiSillLombard}
C.-R. Ji, A. Sill and R. Lombard-Nelsen, {\em Phys. Rev.} {\bf D36},
165 (1987).

\bibitem{JiAmiri}
C.-R. Ji and F. Amiri, {\em Phys. Rev.} {\bf D42}, 3764 (1990).

\bibitem{Davies}
C. T. H. Davies {\it et. al.}, {\em Phys. Rev.} {\bf D52}, 6519
(1995).

\bibitem{BJPR}
S. J. Brodsky, C.-R. Ji, A. Pang, and D. Robertson, SLAC-PUB-7473,
(1997).

\bibitem{Braaten}
E. Braaten and S.-M. Tse, {\em Phys. Rev.}  {\bf D35}, 2255 (1987).

\bibitem{DittesRadyushkin}
F. M. Dittes and A. V. Radyushkin, {\em Sov. J. Nucl. Phys.} {\bf 34},
293 (1981); {\em Phys. Lett.} {\bf 134B}, 359 (1984).

\bibitem{Field}
R. D. Field, R. Gupta, S. Otto and L. Chang, {\em Nucl. Phys.} {\bf
B186}, 429 (1981).


\bibitem{PetronzioParisi}
G. Parisi and R. Petronzio, {\em Phys. Lett.}  {\bf 95B}, 51 (1980).

\bibitem{MattinglyStevenson}
A. C. Mattingly and P. M. Stevenson, {\em Phys. Rev.} {\bf D49}, 437
(1994).

\bibitem{Gribov}
V. N. Gribov, Lund Report No. LU-TP 91-7, 1991 (unpublished).

\bibitem{Zerwas}
K. D. Born, E. Laermann, R. Sommer, P. M. Zerwas, and T. F. Walsh,
{\em Phys. Lett.} {\bf 329B}, 325 (1994).

\bibitem{Cornwall}
J. M. Cornwall, {\em Phys. Rev.} {\bf D26}, 1453 (1982).

\bibitem{DonnachieLandshoff}
A. Donnachie and P. V. Landshoff, {\em Nucl. Phys.} {\bf B311}, 509
(1989).

\bibitem{GayDucatietal}
M. Gay Ducati, F. Halzen and A. A. Natale, {\em Phys. Rev.} {\bf D48},
2324 (1993).

\bibitem{Sloan}
A. X. El-Khadra, G. Hockney, A. Kronfeld and P. B.  Mackenzie, {\em
Phys. Rev. Lett.} {\bf 69}, 729 (1992).

\bibitem{ShirkovMikhailov}
D. V. Shirkov and S. V. Mikhailov, {\em Z. Phys.} {\bf C63}, 463
(1994).

\bibitem{Braun}
V. M. Braun, ``QCD renormalons and higher twist effects,'' {\tt
hep-ph/9505317}.

\bibitem{BrodskyKuhnHoangTuebner}
S. J. Brodsky, A. H. Hoang, J. H. Kuhn and T. Teubner, {\em
Phys. Lett.} {\bf 359B}, 355 (1995).

\bibitem{LewellynIsgur}
N. Isgur and C. H. Lewellyn-Smith, {\em Phys. Rev. Lett.} {\bf 52},
1080 (1984); {\em Phys. Lett.} {\bf 217B}, 535 (1989); {\em
Nucl. Phys.} {\bf B317}, 526 (1989).

\bibitem{Radyushkin}
A. V. Radyushkin, {\em Acta Phys.  Polon.} {\bf B26}, 2067 (1995).

\bibitem{Ong}
S. Ong, {\em Phys. Rev.} {\bf D52}, 3111 (1995).

\bibitem{Kroll}
P. Kroll and M. Raulfs, {\em Phys. Lett.}  {\bf 387B}, 848 (1996).

\bibitem{CZ}
V. L. Chernyak and A. R. Zhitnitsky, {\em Phys. Rep.}  {\bf 112}, 173
(1984).


\end{thebibliography}
\end{document}